\documentclass{elsart}

\usepackage{latexsym}
\usepackage{epsfig}
\usepackage{amssymb}
\usepackage{graphicx}
\begin{document}

\begin{frontmatter}



\title{A fast no-rejection algorithm \\ for the Category Game}


\author[isi]{Francesca Tria},
\author[isi]{Animesh Mukherjee},
\author[spain]{Andrea Baronchelli}, 
\author[roma1,roma2]{Andrea Puglisi} \and
\author[roma1,isi,roma2]{Vittorio Loreto}

\address[isi]{Institute for Scientific Interchange (ISI),\\ Viale Settimio Severo 65, 10133 Torino, Italy}
\address[spain]{Departament de Fisica i Enginyeria Nuclear, Universitat Politecnica de Catalunya, Campus Nord B4, 08034 Barcelona, Spain}
\address[roma1]{Dipartimento di Fisica, Sapienza Universit\`a di Roma,\\ Piazzale Aldo Moro 5, 00185 Roma, Italy}
\address[roma2]{CNR-ISC Piazzale Aldo Moro 5, 00185 Roma, Italy}

\begin{abstract}
The Category Game is a multi-agent model that accounts for the emergence of shared categorization patterns in a population of interacting individuals. In the framework of the model, linguistic categories appear as long lived consensus states that are constantly reshaped and re-negotiated by the communicating individuals. It is therefore crucial to investigate the long time behavior to gain a clear understanding of the dynamics. However, it turns out that the evolution of the emerging category system is so slow, already for small populations, that such an analysis has remained so far impossible. Here, we introduce a fast no-rejection algorithm for the Category Game that disentangles the physical simulation time from the CPU time, thus opening the way for thorough analysis of the model. We verify that the new algorithm is equivalent to the old one in terms of the emerging phenomenology and we quantify the CPU performances of the two algorithms, pointing out the neat advantages offered by the no-rejection one. This technical advance has already opened the way to new investigations of the model, thus helping to shed light on the fundamental issue of categorization. 

\end{abstract}

\end{frontmatter}

\section{Introduction}


The Category Game (CG) is a computational model in which a population of individuals co-evolve their own system of symbols and meanings by playing elementary language games~\cite{cg_pnas}. It has been introduced to investigate how categorization can emerge from scratch in a group of individuals who interact in a pairwise way without any central coordination. The reference problem is color categorization, which is a central issue both in linguistics~\cite{taylor2003lc} and in cognitive science~\cite{deacon1998symbolic,gardner1985msn,lakoff-women}. Color naming represents in fact a fundamental access point to human cognition, and at the same time provides important clues on language evolution. The evolution of English color categories constitutes an excellent example. English color terms exhibited a gradual semantic shift from largely brightness color concepts (Old English) to almost exclusively hue concepts (Middle English)~\cite{casson1997}. The World Color Survey, moreover, showed that color systems across language are not random \cite{berlinkay,kay2003rqc}, but rather exhibit certain statistical regularities, thus opening the way to a revolution in cognitive science~\cite{lakoff-women,murphy2004bbc}.

The main point of interest of the CG is that it is able to reproduce qualitatively and, most remarkably, quantitatively the empirical data gathered in the WCS~\cite{pnas_wcs_2010}. It also differs from the other models defined to address similar issues~\cite{cg_pnas,pnas_wcs_2010,Belpaeme_Bleys2005,bleys2009grounded,dowman2007ect,jameson2009evolutionary,jameson2009evolutionary2,komarova2008pha,komarova2007emc,steels2005cpg} in that it accounts for the categorization of a genuinely continuous perceptual channel and it describes a categorization pattern as a continuously evolving metastable state on which the population shares a temporary consensus \cite{cg_pnas,mukherjee10}. The latter characteristic is intriguing and underlies the existence of a new framework to address the puzzling problem of language change, which turns out to be at the same time propelled by the interaction among the speakers and impeded by the need of these speakers to understand each other~\cite{mukherjee10}. The presence of this sort of frustration renders the dynamics of the model so slow that it has been so far impossible to investigate and quantify properly the details of the long time behavior, even for small population sizes.

Here we present a fast algorithm suitable for studying the CG dynamics over large timescales and for moderately large population sizes. The algorithm is, in spirit, similar to those suggested for accelerating Monte Carlo simulations (see~\cite{kolesik2003,novotny95} and also~\cite{binder97} for other examples), where the key ingredient is to avoid rejection steps (hence the ``no-rejection" tag). Of course, however, the dynamics we are referring to is substantially different, so new methods had to be developed in order to tackle the specific aspects of the model under consideration. The rest of the paper is organized as follows. In section~\ref{cg_describe}, we outline a detailed description of the CG model. The next section presents the fast algorithm suitable for investigating the long-time CG dynamics. Section~\ref{compare_results} compares the outcomes of the fast algorithm with the original one, showing excellent qualitative as well as quantitative agreement. In this context, a detailed investigation of a set of relevant observables is also performed, shedding new light on the ``microscopic'' origin of the ``macroscopic'' behavior of the system.
 In section~\ref{compare_phystime} we show the computational complexity of the proposed fast algorithm, compared with the one of the original model. We conclude in section~\ref{conc} by summarizing our contributions and pointing to possible future applications of this method.

\section{The Category Game model}\label{cg_describe}

The Category Game model \cite{cg_pnas} is crafted to examine how a population of interacting individuals can develop through a series of language games~\cite{wittgenstein53english} a shared form-meaning repertoire from scratch without any pre-existing categorization.  We consider a population of $N$ artificial agents each of them having, without any loss of generality, a one-dimensional continuous perceptual space spanning the $[0, 1)$ interval. A categorization can be identified as a partition of this space into discrete sub-intervals which we shall denote from now onwards as perceptual categories. Each agent has a dynamical inventory of form-meaning associations linking the perceptual categories (meanings) to words (forms). The perceptual categories as well as the words associated to them co-evolve over time through a series of simple communication interactions (or ``games'').

In each game, two individuals are randomly selected from the population and one of them is assigned the role of a speaker while the other the role of a hearer. Both the speaker and hearer are presented with a scene of $M \ge 2$\footnote{Without any loss of generality in all our simulations we shall use $M=2$.} stimuli (objects) where each stimulus corresponds to a real number in the $[0, 1)$ interval. By definition, no two stimuli appearing in the same scene can be at a distance closer than $d_\mathrm{min}$ which is the only parameter of the model encoding the finite resolution power of any perception (for instance, the human Just Noticeable Difference).

One of the objects is the {\em topic} of the communication. The task of the speaker is to communicate this to the hearer using the following prescription. The speaker utters a word associated with the topic while the hearer tries to guess its meaning from the word she ``listened". The speaker always checks whether the topic is the unique among the presented stimuli to lie in the corresponding perceptual category. If it is not, i.e., if the two stimuli collide on the same perceptual category, then a new boundary is created in the perceptual space at a location corresponding to the middle of the segment connecting the two stimuli. A new word is invented for each of the resultant two new categories. In addition, both of them inherit all the words corresponding to the old category. This process is termed as {\em discrimination}. Subsequently, the speaker utters the ``most relevant'' name for the category corresponding to the topic where the most relevant name is either the one used in a previous successful communication or the newly invented name in case the category has just been created due to a discrimination. For the hearer, there can be the following possibilities: (a) the hearer does not have any category associated with the name, in which case the game is a failure, (b) there are one or more categories associated with this name in the hearer's inventory. In this case, the hearer randomly chooses one of them. If the category chosen corresponds to that of the topic, the game is a success, otherwise it is a failure.

Depending on the outcome of the game one or both the agents update their repertoires. In case of a failure, the hearer adds the word in her repertoire linked to the category corresponding to the topic. In case of a success, this word becomes the most relevant name for the category corresponding to the topic for both agents and they remove all other competing words from their respective repertoires linked with this category.  Note that if both the speaker and the hearer already have only the successful word in the corresponding category, the inventory of both of them remains unaltered after the game. This situation, as already mentioned in the introduction, corresponds to a rejection step of a Monte Carlo algorithm. The time $t$ of the dynamics is simply measured as the number of games played by the agents.

The CG dynamics results in the emergence of a hierarchical category structure comprising two distinct levels: a basic layer, responsible for the fine discrimination of the perceptual space (i.e., the perceptual categories), and a second shared linguistic layer that groups together perceptions having the same name to guarantee communicative success (linguistic categories). Note that while the number of perceptual categories is tuned by a parameter of the model and can be very large (of the order of $1/d_\mathrm{min}$), the number of linguistic categories turns out to be finite and small, as observed in natural languages (see fig.\ref{fig1}).

\section{A fast algorithm for the Category Game}\label{cg_fast}

The primary goal of developing a fast algorithm is to study the long time dynamics of the model. In the original algorithm accessing such long timescales would be extremely costly because of the freezing of the dynamics into metastable states. In particular, there would be many games in which the two agents (i.e., the speaker and the hearer) would end up making no changes in the configuration of their respective inventories. The time would then increase by $one$ time step. The basic idea behind the current version of the algorithm is to overcome this freezing by imposing that each game produces an {\em outcome}, i.e., a change in one or both the inventories of the two agents that are playing. In this case, the time $t$ has to be properly rescaled to recover the frozen dynamics. Furthermore, the correct statistical frequency of the different games (the playing order of the pairs and the probability to play in a given region) has to be reproduced in the no-rejection version of the model. In order to do so, we need to calculate both the probability that a game produces an {\em outcome} and the individual probabilities of each possible way in which the outcome can be obtained.

In particular, the main steps of the fast algorithm are the following:

\begin{description}

\item [Choosing the pair of players:] this is the most important part of the algorithm since one has to choose a pair such that their game will produce an outcome. Let us define $p_{\mathrm{out}}(i,j)=p_{\mathrm{out}}(j,i)$ as the probability that a game between the two players will produce an outcome. We shall shortly describe the detailed method for computing this probability. For the time being, let us say that each of the $N(N-1)/2$ possible pairs will be extracted according to this probability\footnote{The actual probability with which each couple is extracted reads: $p_{\mathrm{out}}(i,j)/\sum_{i' < j'} p_{\mathrm{out}}(i',j') $.}.

  \vspace{0.5cm}

\item [Choosing the region for placing the topic:] once the pair of players has been chosen, we impose that the game will produce an outcome. As we shall see below, an outcome can follow if the topic falls either in a mismatch region or in a soft match one (both of these terms will be defined shortly). Subsequently, we need to choose the region to place the topic proportional to the corresponding probabilities $p\mathrm{^{MIS}_{out}}(i,j)$ (for the mismatch region) and $p\mathrm{^{SOFT}_{out}}(i,j)$ (for the soft match region)\footnote{Note that $p\mathrm{^{MIS}_{out}}(i,j)+p\mathrm{^{SOFT}_{out}}(i,j) = p_{\mathrm{out}}(i,j)$. If we want to consider the conditional probabilities of having a game in a soft match region or in a mismatch region {\em given} that the game has an outcome, we have to divide the above defined probabilities by $p_{\mathrm{out}}(i,j)$.}.

  \vspace{0.5cm}

\item[Game:] the game is performed in the selected region with the same rules as in the original CG algorithm summarized in section~\ref{cg_describe}.

  \vspace{0.5cm} 

\item[Rescaling time:] at the end of each game, time is increased by a factor $1/p_{\mathrm{out}}(i,j)$.  

\end{description}

Note that, of course, each probability is also a function of time (we have not explicitly displayed this dependence so as to keep the notations as simple as possible; we shall maintain this implicit form from now onwards throughout the rest of the paper).



\subsection{Extracting the no-rejection regions}

In order to calculate the probabilities of interest, we introduce the following definitions: (i) a { \bf match} region for the two playing agents is a region where both the agents have the corresponding linguistic category settled (i.e., with only one label) and a unique label for it; (ii) a {\bf mismatch} region is any interval in $[0,1)$ that is not a match region.  

Let us now consider a game where a speaker-hearer pair and two stimuli are selected\footnote{without any loss of generality we shall consider from now onwards the first stimulus as being the topic of the game.}. Of course, if the topic falls in a mismatch region an outcome is guaranteed. If the topic falls instead in a match region, the situation is more tricky. In this latter event, two cases are possible: either there is the necessity to discriminate (see section~\ref{cg_describe}) for one or both the agents, in which case the game produces an outcome, or the repertoire of both the speaker and the hearer remains unchanged (the outcome is null). To distinguish between these two events, let us refine the definition of a match region in the following way: we will denote a match region to be a {\bf strict match} region if the lengths of the
  corresponding perceptual categories of both the speaker and the
  hearer are shorter than $d_\mathrm{min}$. In this case no discrimination is possible, due to the finite resolution constraint, and the outcome of the game will be surely null.  A match that is not strict shall be called a {\bf soft match}.

\subsubsection{Probability of playing in a soft match region}

Since the soft match definition is based on the length of the underlying perceptual categories, it is natural to express the probability of having a game in that region as the sum of the probabilities of having a game in each of the underlying perceptual categories:

\begin{equation}
p\mathrm{^{SOFT}_{out}}= \sum_{a} p_a\mathrm{^{SOFT}} \,,
\end{equation}

\noindent where the sum is over all the perceptual categories spanning the soft match region and we assume the dependence on the agents $i,j$ to be implicit.

Thus, we have to calculate the probability $p_a\mathrm{^{SOFT}}$ that both the topic and the object fall in the perceptual category $a$ which belongs to a soft perceptual match region and the game produces an outcome. Two events have to be considered simultaneously: (a) the topic falls in the match region, and (b) the object falls in the union\footnote{The union is the region in $[0,1)$ that belongs either to the perceptual category of the speaker or of the hearer.} of the two perceptual categories under consideration (i.e., of the speaker and the hearer) so as to produce a discrimination and, thereby, an outcome. As we recall from the above, the two events considered here are not independent but correlated through $d_\mathrm{min}$. Consequently, we need to integrate all the different possibilities of placing the topic and the object maintaining this correlation.  In order to write the expression for $p_a\mathrm{^{SOFT}}$, we define $x^{S}_{l,r}$ as the left and the right boundary, respectively, of the considered perceptual category of the speaker, and correspondingly for the hearer. Further, $x^{\mathrm{min}}_{l,r} = \min(x^{S}_{l,r} , x^{H}_{l,r})$, $x^{\mathrm{max}}_{l,r} = \max(x^{S}_{l,r} , x^{H}_{l,r})$, $X^{\mathrm{max}}=\max{(x^{\mathrm{max}}_{l} ,x^{\mathrm{min}}_{l} + d_\mathrm{min})}$ and $X^{\mathrm{min}}=\min{ (x^{\mathrm{min}}_{r} , x^{\mathrm{max}}_{r} - d_\mathrm{min})}$. An example illustrating the above terms is shown in fig.~\ref{fig00}. We finally obtain:

\begin{figure}[htp]
\begin{center}
  \includegraphics[angle=270,width=.95\columnwidth]{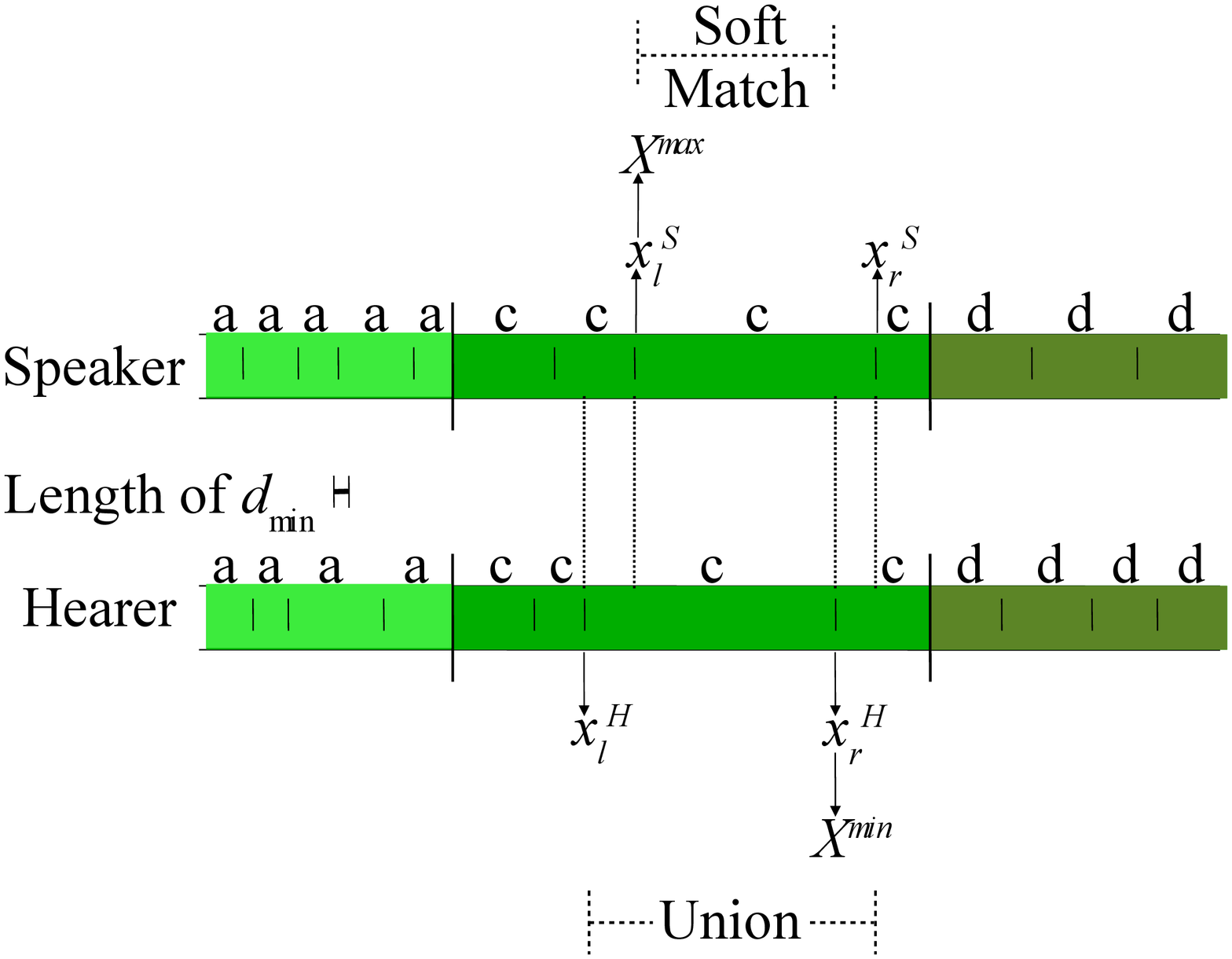}
   \caption{\label{fig00} An example to illustrate the different terms for computing $p_a\mathrm{^{SOFT}}$. According to this figure, (a) $x^{\mathrm{min}}_{l} = \min(x^{S}_{l}, x^{H}_{l}) = x^{H}_{l}$, (b) $x^{\mathrm{min}}_{r} = \min(x^{S}_{r}, x^{H}_{r}) = x^{H}_{r}$, (c) $x^{\mathrm{max}}_{l} = \max(x^{S}_{l}, x^{H}_{l}) = x^{S}_{l}$, (d) $x^{\mathrm{max}}_{r} = \max(x^{S}_{r}, x^{H}_{r}) = x^{S}_{r}$, (e) $X^{\mathrm{max}}=\max(x^{S}_{l}, x^{H}_{l} + d_\mathrm{min}) = x^{S}_{l}$ (taking into account the length of $d_\mathrm{min}$ shown in the figure), and (f) $X^{\mathrm{min}}=\min(x^{H}_{r} , x^{S}_{r} - d_\mathrm{min}) = x^{H}_{r}$ (taking into account the length of $d_\mathrm{min}$ shown in the figure).}
    \end{center}
\end{figure}


\begin{eqnarray}\label{eq:pmatch}
p_a\mathrm{^{SOFT}}&=&\int_{X^{\mathrm{max}}}^{x^{\mathrm{min}}_{r} } dy \int_{x^{\mathrm{min}}_{l} }^{y-d_\mathrm{min}} dz +\int_{x^{\mathrm{max}}_{l}}^{ X^{\mathrm{min}}} dy \int_{y+d_\mathrm{min}}^{x^{\mathrm{max}}_{r}  } dz = \nonumber \\
&=&\frac{(x^{\mathrm{min}}_{r})^2-(X^{\mathrm{max}})^2 }{2}  -(x^{\mathrm{min}}_{r} -X^{\mathrm{max}})(x^{\mathrm{min}}_{l} +d_\mathrm{min}) \nonumber \\ 
&-&\frac{(X^{\mathrm{min}})^2-(x^{\mathrm{max}}_{l})^2}{2} +
 (X^{\mathrm{min}}-x^{\mathrm{max}}_{l} )(x^{\mathrm{max}}_{r} -d_\mathrm{min}) \,.
\end{eqnarray}

\subsubsection{Probability of playing in a mismatch region}


In order to obtain the probability of playing in 
a mismatch region, we have to consider once again  the correlation due to the minimal distance $d_\mathrm{min}$  in the extraction of the topic and the object.
The  probability of producing  an outcome 
by means of a game in a mismatch region equals the probability that the topic falls in there.  In order to calculate this probability, we  
must distinguish the case in which the first extracted object is the topic from the case it is not. 
In the first case, where the topic is the first extracted object, the probability $p_a{\mathrm{^{MIS}_{first}}}$ that it falls in a mismatch region is simply the length of that region.
In the second case, where the topic is the second extracted object, we have to consider the fact that it should be chosen at least $d_\mathrm{min}$ far apart from the first object.
This can be done in a similar way as described in the previous subsection for the calculation of the probability of playing in a soft match region and we shall call this probability $p_a{\mathrm{^{MIS}_{second}}}$.
Since the first and second case occur with equal probability in the original model, we can write:
\begin{eqnarray}
p_a{\mathrm{^{MIS}}}&=&\frac{1}{2}(p_a{\mathrm{^{MIS}_{first}}}+p_a{\mathrm{^{MIS}_{second}}}) \\
p\mathrm{^{MIS}_{out}}&= &\sum_{a} p_a\mathrm{^{MIS} } \,,
\end{eqnarray}
where the sum is over all the perceptual categories spanning the mismatch region.

\section{Comparison of the observables from the two algorithms}\label{compare_results}

In this section, we show that this no-rejection version of the CG algorithm features the same dynamical properties of the original one. We consider all the relevant observables reported in~\cite{cg_pnas} as well as new microscopic observables analyzed here for the first time.

\subsection{``Macroscopic'' observables}

Here we look at (a) the average number of perceptual ($n_{perc}$) and linguistic categories ($n_{ling}$) per individual, (b) the success rate and (c) the perceptual and linguistic overlap across the population as functions of time for the two models (original and fast) we are comparing.  We show simulations for different population sizes and a fixed $d_\mathrm{min} = 0.0143$ which is equal to the average human JND~\cite{pnas_wcs_2010}.

\begin{figure}
\begin{center}
  \includegraphics[width=.55\columnwidth]{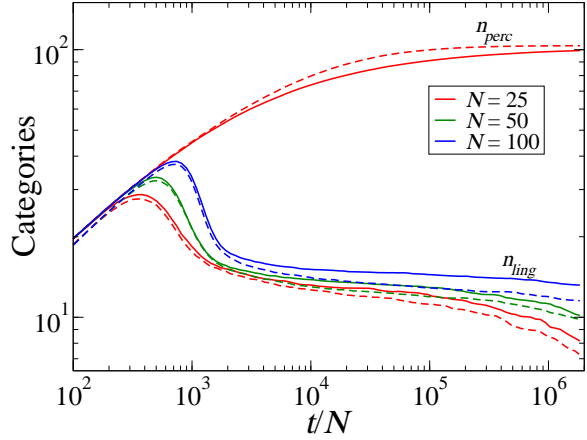}
   \caption{\label{fig1} Perceptual and linguistic categories obtained from the old and the new algorithm for $N$ = 25, 50 and 100. Solid lines show results obtained from the old algorithm while broken lines indicate results obtained from the new algorithm. All the results are averaged over 30 samples.}
    \end{center}
\end{figure}

Fig.~\ref{fig1} shows the average number of perceptual\footnote{The average number of perceptual categories remains same across different population sizes as long as the value of $d_\mathrm{min}$ is fixed. Therefore we only show one representative plot for the perceptual categories in fig.~\ref{fig1}.} and linguistic categories per individual obtained from the old and the new algorithms respectively versus the number of games per player (i.e., $t/N$). It is apparent from this figure that the new algorithm is not only able to reproduce the same phenomenology but also the outcomes are very close to what is obtained from the old algorithm.

\begin{figure}
\begin{center}
  \includegraphics[width=0.6\columnwidth]{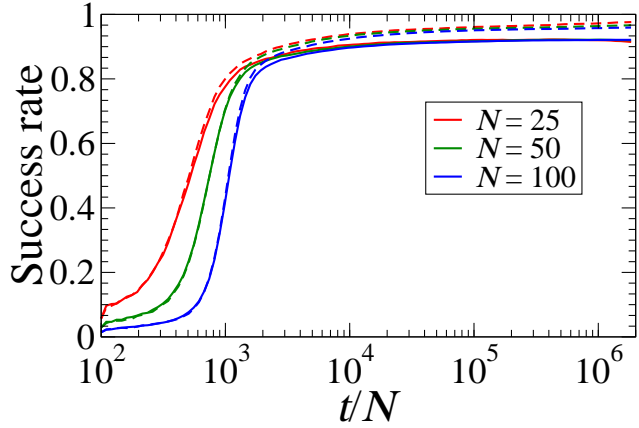}
  \caption{\label{fig2} Success rate obtained from the old and the new algorithm for $N$ = 25, 50 and 100. Solid lines show results obtained from the old algorithm while broken lines indicate results obtained from the new algorithm. All the results are averaged over 30 samples.}
    \end{center}
\end{figure}

Fig.~\ref{fig2} compares the success rate for the two algorithms versus $t/N$. The success rate is measured as the fraction of successful games over sliding time windows. The figure clearly indicates that the results from the two algorithms match (almost) accurately.
\\
\\

The overlap function~\cite{cg_pnas} measures the degree of alignment of the category boundaries of two agents ($i$, $j$) and is defined as follows:

\begin{equation}\label{eq:ov}
  O = \frac{2}{N(N-1)} \sum_{i<j} O_{ij}\textrm{ with } 
  O_{ij} = \frac{2\sum_{c_i^{j}} {(l_{c_i^{j}})}^2}
  {\sum_{c_i} {(l_{c_i})}^2 + \sum_{c_{j}} {(l_{c_{j}})}^2}
\end{equation}

\noindent where $l_c$ is the width of the category $c$, $c_{i}$ is one of the categories of the $i^{th}$ agent and $c_i^{j}$ is the generic category of the intersection set containing all of the boundaries of the agents $i$ and $j$. The function $O_{ij}$ returns a value proportional to the degree of alignment of the two category inventories reaching its maximum unitary value when they are perfectly aligned. The linguistic overlap is defined as in eq.~\ref{eq:ov} however, only considering the boundaries between categories with different most relevant names.

\begin{figure}[htp] \begin{center}
  \includegraphics[width=0.6\columnwidth]{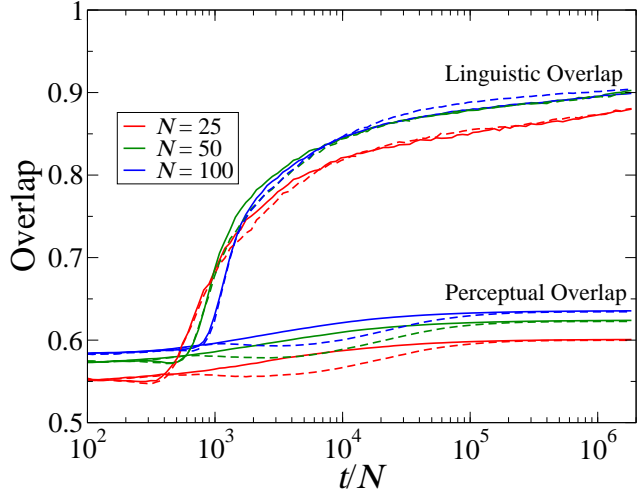}
  \caption{\label{fig3} Perceptual and linguistic overlap obtained from the old and the new algorithm for $N$ = 25, 50 and 100. Solid lines show results obtained from the old algorithm while broken lines indicate results obtained from the new algorithm. All the results are averaged over 30 samples.}
    \end{center}
\end{figure}

Fig.~\ref{fig3} shows the perceptual and linguistic overlap obtained from the old and the new algorithm versus $t/N$. Once again there is an excellent agreement between the results obtained from the two algorithms.

\subsection{``Microscopic'' observables}

\begin{figure}
\begin{center}
  \includegraphics[width=\columnwidth]{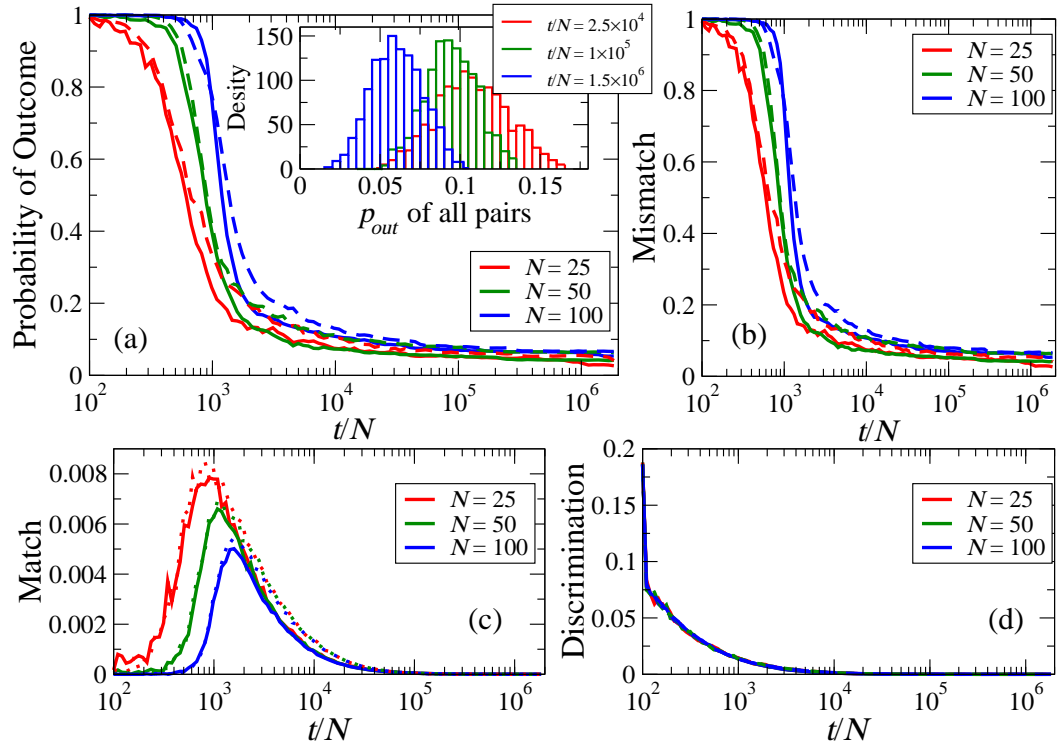}
  \caption{\label{fig0} Microscopic observables. (a) Probability of having an outcome. The solid lines show the average fraction of games collected in time sliding windows versus $t/N$ for the old algorithm. The broken lines show the average values of $p_{\mathrm{out}}$ in the same time sliding windows versus $t/N$ for the new algorithm. The inset shows the (representative) histogram of $p_{\mathrm{out}}$ for each possible pair of agents at three different points in time for $N$ = 50. (b) The fraction of games, collected in time sliding windows, which ended up being played in a mismatch region for the old (solid lines) and the new algorithm (broken lines). (c) The fraction of games, collected in time sliding windows, which ended up being played in a match region for the old (solid lines) and the new algorithm (broken lines). (d) The fraction of games, collected in time sliding windows, where a discrimination process occurred for the old (solid lines) and the new algorithm (broken lines). All our results are for $N$ = 25, 50 and 100 and each of them are averaged over $30$ samples.}
    \end{center}
\end{figure}

As pointed out in the previous sections, one of the main ingredients of the fast algorithm is the computation of the probability that at each time step a game between a randomly chosen pair of players together with a randomly selected topic in the interval $[0,1)$ would produce a non-null outcome.  Here we compare the behavior of the original and the fast algorithm with respect to this property.  In particular, in fig.~\ref{fig0}(a) we show for the original algorithm the fraction of games, collected in time sliding windows, that produce a non-null outcome, while for the no-rejection algorithm the probability $p_{\mathrm{out}}(i,j)$ of the selected pair $(i,j)$ averaged over the same time windows.  In the inset, we also show the (representative) histogram of $p_{\mathrm{out}}$ for each possible pair of agents at three different points in time for a specific population size.  It is evident that the histogram is strongly peaked, which means that $p_{\mathrm{out}}(i,j)$ is roughly similar for all pairs $(i,j)$ thereby, allowing for an effective random choice over the pairs without altering the dynamics.

Subsequently, in fig.~\ref{fig0}(b), (c) and (d) we respectively compare the two algorithms in terms of the fraction of games, collected in time sliding windows, which ended up being played in a mismatch region, or in a match region, or where a discrimination process occurred. Once again, the results exhibit a remarkable qualitative as well as quantitative agreement. 



\section{Comparison of the computational complexity of the two algorithms}\label{compare_phystime}

In this section, we compare the computational complexity of the two algorithms, i.e., we compare the computer time\footnote{We use the in-built {\bf clock()} function of the GNU C library to estimate the value of $t_c$.} ($t_c$) in seconds required to complete a specific number of games per player for the old and the new algorithms. In addition, we give more extensive results for the computational complexity of the fast algorithm.

\begin{figure} \begin{center}
  \includegraphics[width=0.8\columnwidth]{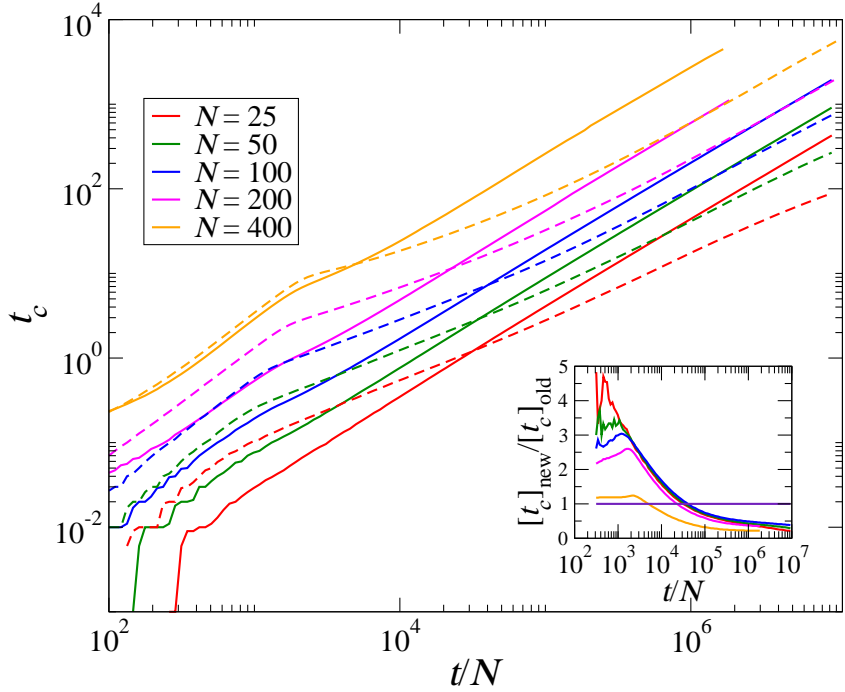}
  \caption{\label{fig4} Computer time ($t_c$) in seconds required to complete a specific number of games per player by the old and the new algorithm for $N$ = 25, 50, 100, 200 and 400. Solid lines show the computer time required by the old algorithm while broken lines indicate the computer time required by the new algorithm. The results are obtained on the ISI cluster with the following node specification: (a) Processor -- INTEL XEON E5405 2.00GHz, (b) Cache -- 6144KB, (c) ISA: 64-bit and (d) RAM -- 8GB. The inset shows the ratio of $t_c$ for the new algorithm ($[t_c]_\mathrm{new}$) to the $t_c$ for the old algorithm ($[t_c]_\mathrm{old}$) for the same population sizes as in the main figure.}
    \end{center}
\end{figure}

Fig.~\ref{fig4} compares the computer time required to complete a specific number of games per player by the old and the new algorithm. Note that during the initial games the old algorithm takes lesser computer time than the new one. During this phase, almost all games produce an outcome and therefore the additional calculation of the probabilities required for the fast algorithm is not advantageous. However, as soon as the dynamics gets trapped in metastable states (at the onset of the plateau region in the number of linguistic categories curve in fig.~\ref{fig1}) the calculation of the probabilities turns out to be very advantageous and the new algorithm has a much higher velocity than the old one. Note that for larger population sizes ($N$ = 200 and 400) we could manage to reach only a lower value of $t/N$ within a reasonable $t_c$ for the old algorithm. In the inset of fig.~\ref{fig4} the ratio of the two algorithms' execution time is shown, to better appreciate the advantage of the no-rejection one.

\begin{figure}[!t]
\begin{center}
  \includegraphics[width=0.9\columnwidth]{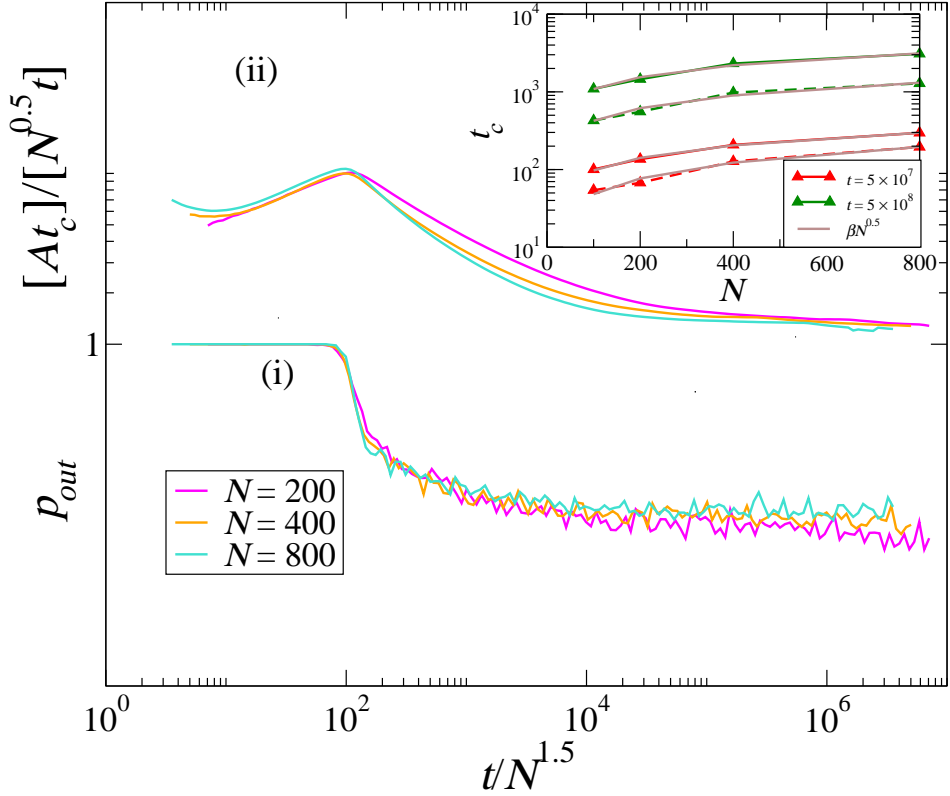}
  \caption{\label{fig5} The scaling of $t_c$ with the population size for the new algorithm. (i) $p_\mathrm{out}$ versus the rescaled number of games. $t$ is rescaled by $N^{3/2}$ in order to collapse the curves. (ii) Rescaled $t_c$ versus the rescaled number of games (i.e., $t/N^{3/2}$ as in (i)). $t_c$ is rescaled by $t\sqrt{N}$ and then the entire factor is multiplied by a large constant $A$ ($\sim 2 \times 10^7$) in order to present a better visualization of the plots within the same figure as of $p_\mathrm{out}$. $N = 200, 400$ and $800$.  The results are obtained on the ISI cluster with the following node specification: (a) Processor -- INTEL XEON E5405 2.00GHz, (b) Cache -- 6144KB, (c) ISA: 64-bit and (d) RAM -- 8GB. The inset shows the amount of $t_c$ required to complete $t = 5 \times 10^7, 5 \times 10^8$ steps for the old (continuous lines) and the new (dashed lines) algorithm for different population sizes. All the four curves can be fitted with a function of the form: $f(N) = \beta\sqrt{N}$. For the old algorithm $\beta = 8.25$ and $104.92$ while for the new algorithm $\beta = 2.16$ and $34.66$ respectively for $t = 5 \times 10^7$ and $t = 5 \times 10^8$.}
    \end{center}
\end{figure}

In fig.~\ref{fig5}, we present a further analysis of the scaling of $t_c$ with the population size for the no-rejection algorithm. In particular, we report two different quantities in this figure for $N$ = 200, 400 and 800:\\
\\
(i) $p_\mathrm{out}$ versus the rescaled number of games. Note that one needs to rescale $t$ by $N^{3/2}$ to collapse the curves. This dependence of the time of the dynamics on the population size has also been recovered in several other cases elsewhere~\cite{mukherjee10,baronchelli_ng_first} and usually indicates the time scaling to reach a consensus in the population. Clearly, the rate at which the values of $p_\mathrm{out}$ drop decreases with increasing $t$.
\\
(ii) the rescaled $t_c$ versus the rescaled number of games (i.e., $t/N^{3/2}$ as in case (i)). In order to collapse these curves, especially in the ``large" $t$ regime (featuring the long-time dynamics) one has to rescale $t_c$ by $ t \sqrt{N}$. In addition the entire factor is multiplied by a large constant $A$ ($\sim 2 \times 10^7$) in order to present a better visualization of the plots within the same figure as of $p_\mathrm{out}$.

The most important observation is that the long-time behavior is exactly similar to that of $p_\mathrm{out}$ which indicates that the computer time required is largely determined by the probability of outcomes. The inset of the same figure shows the amount of $t_c$ required to complete $t = 5 \times 10^7, 5 \times 10^8$ steps for the old and the new algorithm for different population sizes. In all the four cases, the curves can be nicely fitted using a function of the form: $f(N) = \beta\sqrt{N}$. This observation once again confirms the dependence: $t_c \propto \sqrt{N}$. An important point to note is that the pre-factor $\beta$ is significantly lower for the new algorithm as compared to the old one. The dependence of $t_c$ on $\sqrt{N}$ can be attributed to the time required by the different processes of the model (e.g., discrimination, inventory updates etc.). A detailed analysis of this dependence is out of the scope of the current paper and shall be presented elsewhere.

\section{Conclusion}\label{conc}

In this paper we have introduced a no-rejection algorithm developed to study the long time behavior of the Category Game model.  The original model~\cite{cg_pnas} approaches one of the most important problems in linguistics -- the emergence of linguistic categories -- and was shown to reproduce both qualitatively and quantitatively experimental results reported in the WCS~\cite{pnas_wcs_2010}.  The two main innovative aspects of the model, with respect to previously proposed ones, are (i) the dynamical emergence of a discretization from a continuous perceptual space and (ii) the representation of the present category system as a long lasting metastable state rather then an attractor of the dynamics.  The last property has triggered the need for suitable methods to achieve the long time dynamics of the model.  In particular, the observed dynamics was such that games which bring a modification of the state of the agents resulted progressively more rare. The analysis of the behavior of the system for suitably large time and population sizes was for that reason practically impossible.  The no-rejection algorithm presented here has been crucial to access the long time dynamical properties of CG, characterized by metastability and aging~\cite{mukherjee10}, and thus to shed light on so far unexplored part of the model, helping to clarify the nature of the consensus states that are established during the CG dynamics, with important consequences for the understanding of such phenomena as language change and language evolution.  The no-rejection model we propose, despite being {\em ad hoc} for the CG, presents principles and methods that can be generalized for different agent based models and we believe that that could trigger computationally less expensive methods suitable to investigate social phenomena.

 
\bibliographystyle{unsrt} 
\bibliography{els_bib}

\end{document}